\documentstyle[sprocl,epsf]{article}

\def\draftversion{N}                        
\def\note[#1]#2{\message{(#1)}\if\draftversion Y{\noindent\em #2}\fi}
\if \draftversion Y
\fi
\def\spose#1{\hbox to 0pt{#1\hss}}
\def\ltapprox{\mathrel{\spose{\lower 3pt\hbox{$\mathchar"218$}}
 \raise 2.0pt\hbox{$\mathchar"13C$}}}
\def\gtapprox{\mathrel{\spose{\lower 3pt\hbox{$\mathchar"218$}}
 \raise 2.0pt\hbox{$\mathchar"13E$}}}
\def\inapprox{\mathrel{\spose{\lower 3pt\hbox{$\mathchar"218$}}
 \raise 2.0pt\hbox{$\mathchar"232$}}}

\def\slash{\!\!\!\!/\,}
\def\Dslash{{D}\slash}

\begin{document}

\vskip -3.5cm
\rightline{ZiF-MS-02/96}
\rightline{FSU-SCRI-96C-122}
\rightline{UK/96-14}
\vskip 1.0cm

\title{REDUCING DISCRETIZATION ERRORS IN LATTICE QCD SPECTROSCOPY\footnote{Talk
given by UMH at the Conference ``Multi-Scale Phenomena and Their
Simulation'', September 30 -- October 5, 1996, Bielefeld, Germany. To
appear in the proceedings.}}

\author{Sara~Collins}

\address{Dept. of Physics, University of Glasgow,
         Glasgow G12 8QQ, Scotland}

\author{Robert~G.~Edwards, Urs~M.~Heller}

\address{SCRI, Florida State University, Tallahassee, FL 32306--4052, USA}

\author{John~Sloan}

\address{Dept. of Physics, University of Kentucky,
         Lexington, KY 40506--0055, USA}

\maketitle
\abstracts{
The improved Wilson quark action -- the clover action -- is constructed
to have smaller discretization errors than the normal Wilson quark
action. We test this in a quenched spectroscopy computation on 6 lattice
ensembles with spacings from $0.15$ to $0.43$ fm. To ensure that the
dominant scaling violations come from the fermions we use an
${\cal O}(a^2)$ improved 6-link $SU(3)$ pure gauge action. We
find evidence that fermionic scaling violations are consistent with
${\cal O}(a^2)$ for clover fermions and ${\cal O}(a)$ with a nonnegligible
${\cal O}(a^2)$ term for standard Wilson fermions. This latter mixed
ansatz makes a reliable continuum extrapolation problematic for Wilson
fermions. For clover fermions, on the other hand, we obtain accurate
predictions for hadron masses in quenched continuum QCD. We find that
the slopes of the scaling violations are roughly 200 MeV for both Wilson
and clover fermions.}

\section{Introduction}

The goal of lattice QCD spectroscopy computations is, of course, to make
mass predictions for {\it continuum} QCD. This requires in the end an
extrapolation from results at finite lattice spacing $a$ to the
continuum limit, $a=0$. We call the deviations of the finite lattice
spacing results from their continuum limit value discretization errors.
For Wilson fermions, the leading discretization errors are known to be
of order ${\cal O}(a)$. In other words, relative discretization errors
are expected to be about $a\Lambda_{\rm QCD}$, with $\Lambda_{\rm QCD}
\approx 300 - 500$ MeV a typical QCD scale. In a typical simulation
today, say at $\beta=6$, where $a^{-1} \simeq 2$ GeV, we therefore
expect discretization errors for computations with Wilson fermions of
15 -- 25\%. Even at the smallest lattice spacing reached to date, the
errors are expected to be $\gtapprox 10\%$.

Decreasing the lattice spacing to make the discretization errors smaller
is a costly enterprise. Just the cost to create pure gauge configurations
of constant physical volume grows like $a^{-5}$, and the cost to perform
quenched spectroscopy computations at constant pion to rho mass ratio on
these configurations grows probably even faster. For full QCD simulations
the situation is even worse.\cite{Philippe,Karl} Therefore it seems
worthwhile to `improve' the action so as to reduce the {\it order} of
the discretization errors from ${\cal O}(a)$ to ${\cal O}(a^2)$, as
will be described in this talk, or even to higher order.\cite{Tim}

\section{The Symanzik improvement program}

Symanzik has first proposed that lattice effects, leading to
discretization errors, can be eliminated, order by order in $a$, by
including irrelevant, higher dimension operators into the lattice
action.\cite{Symanzik} For example, the simplest pure gauge lattice
action, the so-called Wilson action
\begin{displaymath}
  S = \frac{\beta}{N} \sum_p {\rm Re Tr}(1 - U_p)
\end{displaymath}
becomes, as $a \to 0$
\begin{displaymath}
  S \rightarrow \frac{1}{2} a^4 \sum_{x,\mu,\nu} \left\{ {\rm Tr}
(F^2_{\mu \nu}) - \frac{1}{12} a^2 {\rm Tr}(D_\mu F_{\mu \nu} D_\mu
F_{\mu \nu}) + {\cal O}(a^4) \right\} .
\end{displaymath}
The ${\cal O}(a^2)$ contribution can be canceled by adding a
$2\times1$-loop term
\begin{displaymath}
  S = \frac{\beta}{N} \left\{ \frac{5}{3} \sum_p {\rm Re Tr}(1 - U_p)
  - \frac{1}{12} \sum_{2\times1} {\rm Re Tr}(1 - U_{2\times1}) \right\} .
\end{displaymath}
Loop corrections (quantum effects) modify the coefficients, $5/3$ and
$-1/12$, and require
an additional dimension 6 operator, $\sum_{\mu,\nu,\rho} {\rm Tr}
(D_\mu F_{\nu \rho} D_\mu F_{\nu \rho})$, represented on the lattice by
a 6-link generalized parallelogram along the edges of a simple lattice
cube.\cite{Luscher_85}

We have used this one-loop Symanzik-improved pure gauge action (see
ref.~\citelow{Luscher_85} for details) with
tadpole-improved coefficients as described in~\citelow{Alford_95} to
generate our gauge configuration ensembles. Using this improved gauge
action ensures that the discretization effects, which are the subject
of our study, are dominated by the fermions even with the improved
fermions.

Wilson fermions with the commonly used normalization are given by the action
\begin{eqnarray}
  S_W &=& 2 \kappa \sum_x \bar \psi \left[ \Dslash - \frac{1}{2} \Delta
  + m \right] \psi \nonumber \\
  &=& \sum_x \bar \psi(x) \psi(x)
  + \kappa \sum_{x,\mu} \bar \psi(x) \left[ \left( \gamma_\mu
  - 1 \right) U_\mu(x) \psi(x+\mu) \right. \nonumber \\
  && \left. - \left( \gamma_\mu + 1 \right)
  U^\dagger_\mu(x-\mu) \psi(x-\mu) \right] 
  \label{eq:S_W}
\end{eqnarray}
where $\kappa = 1/(8+2m)$.
The dimension 5 (irrelevant) Wilson term $\bar \psi \Delta \psi$
was introduced to lift the doublers,
present with naive fermions on the edges of the Brillouin zone. But this
term leads to ${\cal O}(a)$ discretization errors for the Wilson
fermions.

In the spirit of the Symanzik improvement program, the ${\cal O}(a)$
effects can be canceled by including another dimension 5 operator
\begin{equation}
  - c_{SW} \frac{\kappa}{2} \psi(x) \sigma_{\mu \nu} i F_{\mu \nu}(x)
  \psi(x)
  \label{eq:SW_term}
\end{equation}
into the action Eq.~(\ref{eq:S_W}) as noted by Sheikholeslami and
Wohlert.\cite{SW_85} On the lattice $F_{\mu \nu}$ is represented by a
clover leaf of plaquettes around site $x$. For this reason these
improved fermions are often referred to as clover fermions. The clover
coefficient $c_{SW}$ is equal to 1 at tree level and equal to $1/u^3_0$,
where $u_0^4 = \langle {\rm Tr} U_p/3 \rangle$, with tadpole
improvement. It has been shown that, with Wilson pure gauge action, the
one-loop coefficient is dominated by the tadpole contribution.\cite{Naik_93}
With Wilson pure gauge action the clover coefficient has
recently also been determined nonperturbatively.\cite{Alpha_96} Both
the one-loop computation and the nonperturbative determination do not
apply to the tadpole-improved 1-loop Symanzik pure gauge action used by
us. We therefore used the tadpole-improved tree-level clover coefficient,
$c_{SW} = 1/u^3_0$.

\section{Comparing Clover with Wilson fermion spectroscopy}

To test whether clover fermions have indeed reduced discretization
errors as compared to Wilson fermions we made an extensive quenched
spectroscopy computation. We generated 6 ensembles of pure gauge
configurations with the tadpole-improved 1-loop Symanzik pure gauge
action with lattice spacings, as measured from the string tension with
$\sqrt{\sigma} = 440$ MeV, of $0.16$, $0.18$, $0.21$, $0.26$, $0.34$, and
$0.42$ fm. All lattices have size $16^3\times32$ and hence our smallest
system has spatial size $L_s \simeq 2.56$ fm. We therefore expect finite
volume effects to be very small. Except for this smallest system we used
the method of `$Z(3)$' fermion sources\cite{Butler_94} to increase the
statistics at no extra computational cost.

For hadron measurements, we used correlated multi-state fits to
multiple correlation functions as discussed in \citelow{Collins_95}. We
computed quark propagators in Coulomb gauge at several $\kappa$ values,
with $m_\pi/m_\rho \gtapprox 0.5$, on each configuration ensemble. We used
two gaussian source smearing functions with smeared and local sinks.
The hadron masses were then extrapolated to the physical $m_\pi/m_\rho$
ratio using correlated chiral fits
\begin{equation}
  m_H =   c_0 + c_2 m^2_\pi + c_3 m^3_\pi  ~~~~~{\rm for} ~~ H = \rho, N,
  \Delta .
  \label{eq:chiral}
\end{equation}
We actually used a cubic fit only for the clover data at the largest
lattice spacing.

\begin{figure}
  \begin{center}
  \vskip 10mm
  \leavevmode
  \epsfxsize=110mm
  \epsfbox[90 80 540 410]{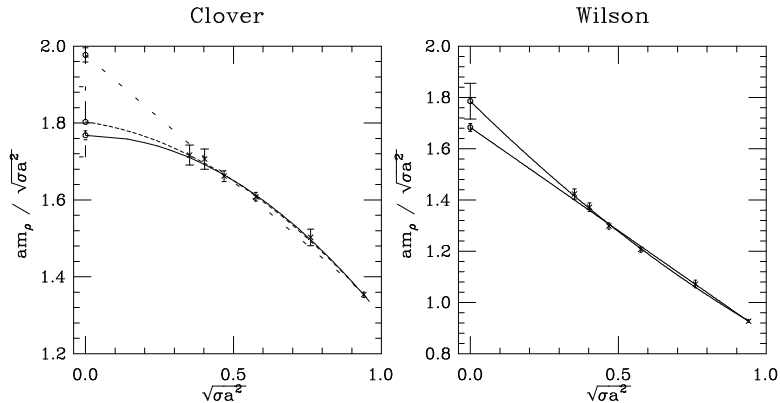}
  \end{center}
  \vskip -40mm
  \caption{Scaling plot of $m_\rho/\protect\sqrt{\sigma}$ versus the lattice
           spacing as measured in units of $\protect\sqrt{\sigma}$.}
  \label{fig:rho_both}
\end{figure}

In Fig.~\ref{fig:rho_both} we show the ratio $m_\rho/\sqrt{\sigma}$
versus the lattice spacing for both clover and Wilson fermions. Also
shown in the figure are scaling fits
\begin{equation}
  R(a) = R(0) ~ \left[ 1 + \mu_1 a + (\mu_2 a)^2 \right]
  \label{eq:scal_fit}
\end{equation}
to the dimensionless ratio. We call the scaling fit linear, when $\mu_2
= 0$, quadratic, when $\mu_1 = 0$, and mixed when both coefficients are
allowed to vary. The confidence level for the different fits are listed
in Table~\ref{tab:Q} and in Table~\ref{tab:slopes} we give the scaling
slopes for the rho, nucleon and delta for linear and quadratic scaling
fits for Wilson and clover fermions, respectively.

\begin{table}
  \begin{center}
    \caption{Confidence level, $Q$, of scaling fits,
             Eq.~(\protect\ref{eq:scal_fit}), to the ratio $R =
             m_\rho/\protect\sqrt{\sigma}$.}
    \vspace{5mm}
    \tabcolsep 4pt
    \begin{tabular}{|c|ccc|}
      \hline
      $Q$ & linear & quadratic & mixed \\ \hline
      Wilson & 0.45 & $\sim 10^{-5}$ & 0.69 \\
      Clover & 0.64 & 0.93 & 0.96 \\
      \hline
    \end{tabular}
    \label{tab:Q}
  \end{center}
\end{table}

\begin{table}
  \begin{center}
    \caption{Scaling slopes $\mu_1$ for Wilson fermions (linear scaling
             fit) and $\mu_2$ for clover fermion (quadratic scaling fit).}
    \vspace{5mm}
    \tabcolsep 4pt
    \begin{tabular}{|c|cc|}
      \hline
      $\mu$ (MeV) & Wilson & Clover \\ \hline
      $m_\rho$   & 215 & 225 \\
      $m_N$      & 115 & 160 \\
      $m_\Delta$ & 170 & 200 \\
      $J_{K^*}$  &  70 & 190 \\
      \hline
    \end{tabular}
    \label{tab:slopes}
  \end{center}
\end{table}

For Wilson fermions both linear and mixed fits work well, but there is a
significant ${\cal O}(a^2)$ contribution. For the mixed fit to
$m_\rho/\sqrt{\sigma}$ we find $\mu_1 = 280$ and $\mu_2 = 160$ MeV with
opposite signs. For clover fermions we can not exclude the linear
scaling fit. However, the quadratic fit works very well, and in the mixed
fit we obtain a linear coefficient compatible with zero:
$\mu_1 = -49 \pm 68$ MeV. We also find that the quadratic fit for clover
fermions is stable when dropping the points with largest lattice spacing,
while this is not the case for the linear fit for Wilson fermions. We
conclude that our clover results are compatible with ${\cal O}(a^2)$
discretization errors, while for Wilson fermions both ${\cal O}(a)$ {\it
and} ${\cal O}(a^2)$ contributions appear significant. We finally note
that the scaling slopes (see Table~\ref{tab:slopes}) are of reasonable
magnitude, maybe even slightly smaller than the $300$-$500$ MeV expected.

\begin{figure}
  \begin{center}
  \vskip 10mm
  \leavevmode
  \epsfxsize=110mm
  \epsfbox[90 80 540 410]{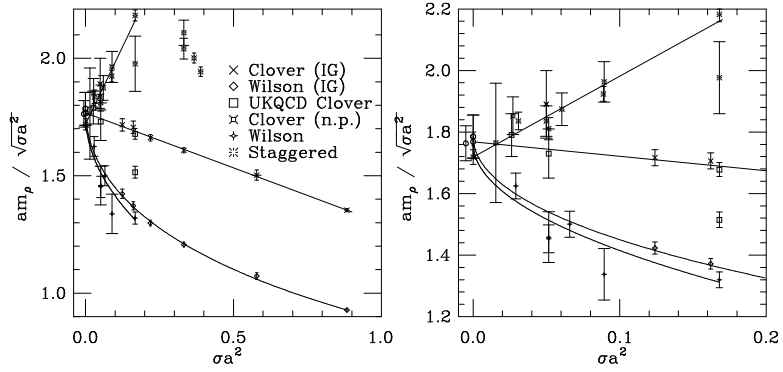}
  \end{center}
  \vskip -40mm
  \caption{Scaling plot of $m_\rho/\protect\sqrt{\sigma}$ versus $a^2$ for a
           collection of recent spectroscopy computations with clover,
           Wilson and staggered fermions. At right is a blowup of
           the small $a$ region.}
  \label{fig:rho_world}
\end{figure}

In Fig.~\ref{fig:rho_world} we compare our results on ``improved glue''
with a collection of recent quenched spectroscopy computations with
clover,\cite{Rho_clov} Wilson \cite{Rho_Wils}and staggered
fermions.\cite{Rho_stag} A quadratic scaling
fit to the staggered fermion data, and a mixed scaling fit to the Wilson
data are also shown. It is satisfying to notice that the continuum
extrapolations all agree within their statistical errors.

\begin{figure}
  \begin{center}
  \vskip 10mm
  \leavevmode
  \epsfxsize=110mm
  \epsfbox[90 80 540 410]{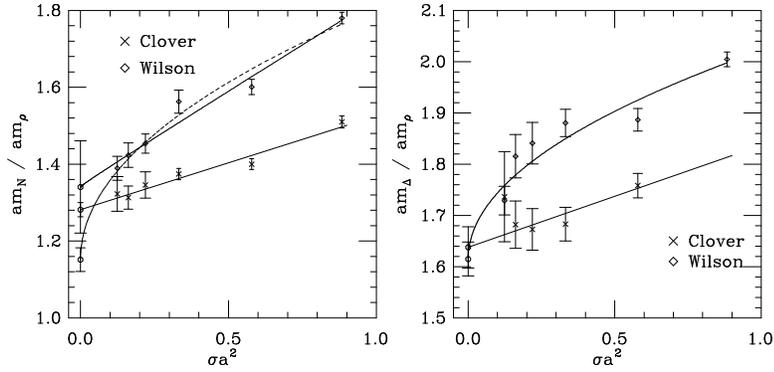}
  \end{center}
  \vskip -40mm
  \caption{Scaling plots of $m_N / m_\rho$ and $m_\Delta / m_\rho$
           versus $a^2$ with quadratic scaling fit for clover fermions.
           For Wilson fermions are shown the linear (dashed) and
           mixed (full) scaling fit for the nucleon, and the linear
           scaling fit for the delta.}
  \label{fig:N_delta_rho_sig}
\end{figure}

In Fig.~\ref{fig:N_delta_rho_sig}, we plot the ratios $m_N / m_\rho$
and $m_\Delta / m_\rho$ assuming ${\cal O}(a^2)$ scaling violations for
clover fermions. For $m_N / m_\rho$ we find a continuum value of
$1.29(2)$ that is consistent with the GF11 result;\cite{Butler_94}
however, it is different than the experimental value of $1.22$. For
Wilson fermions a linear scaling fit gives a value inconsistent with
the clover extrapolation and lower than the experimental value, while
a mixed scaling fit gives a consistent continuum extrapolation, albeit
with large errors. Both scaling fits give similar confidence levels,
$Q$. For $m_\Delta / m_\rho$ we obtain a continuum value of $1.65(4)$
for clover fermions and, using only an ${\cal O}(a)$ ansatz, $1.61(3)$
for Wilson fermions. Both extrapolations agree, within errors, with the
experimental value, $1.60$.

\section{Conclusions}

We have made a comprehensive study of discretization effects for Wilson
and ${\cal O}(a)$ improved clover fermions. This was done in the context
of a quenched spectroscopy calculation on gauge field ensembles with
lattice spacings varying from $0.42$ to $0.16$ fm. The gauge field
configurations were created with an ${\cal O}(a^2)$ improved pure gauge
action, ensuring that the leading discretization errors come from the
fermions. We found the discretization errors for clover fermions to be
compatible with being ${\cal O}(a^2)$, though numerically we could not
exclude small ${\cal O}(a)$ contributions. For Wilson fermions we found
that both ${\cal O}(a)$ and ${\cal O}(a^2)$ contributions are
significant, making a reliable continuum extrapolation difficult. For
the rho mass we found at our largest lattice spacing, $a=0.42$ fm,
scaling violations of $\sim 50\%$ for Wilson and $\sim 25\%$ for clover
fermions. At our smallest lattice spacing, $a=0.16$ fm, they decreased
to $\sim 20\%$ and $\sim 3\%$, respectively. Therefore, for clover
fermions we can make reliable continuum extrapolations with the
extrapolation errors not exceeding a few percent.

We found that the scaling slopes are of reasonable magnitude, $\sim 200$
MeV. This suggests that an expansion in powers of the lattice spacing is
reasonably behaved and converging, encouraging news for attempts of
higher order improvements.\cite{Tim}

\section*{Acknowledgments}
This research was supported by DOE contracts
DE-FG05-85ER250000 and DE-FG05-92ER40742.
UMH would like to express his gratitude for the hospitality
of the Zentrum f\"ur interdisziplin\"are Forschung and the
Fakult\"at f\"ur Physik of the University of Bielefeld
during his stay in Bielefeld.

%
%
%

\end{document}